\documentclass[12pt]{article}

\RequirePackage{cite}
\RequirePackage{times}
\RequirePackage{fullpage}
\RequirePackage{ifthen}

\makeatletter
\def\@cite#1#2{$^{\mbox{\scriptsize #1\if@tempswa , #2\fi}}$}
\makeatother

\newcommand{\spacing}[1]{\renewcommand{\baselinestretch}{#1}\large\normalsize}
\spacing{2}

\def\@maketitle{%
  \newpage\spacing{1}\setlength{\parskip}{12pt}%
    {\Large\bfseries\noindent\sloppy \textsf{\@title} \par}%
    {\noindent\sloppy \@author}%
}

\newenvironment{affiliations}{%
    \setcounter{enumi}{1}%
    \setlength{\parindent}{0in}%
    \slshape\sloppy%
    \begin{list}{\upshape$^{\arabic{enumi}}$}{%
        \usecounter{enumi}%
        \setlength{\leftmargin}{0in}%
        \setlength{\topsep}{0in}%
        \setlength{\labelsep}{0in}%
        \setlength{\labelwidth}{0in}%
        \setlength{\listparindent}{0in}%
        \setlength{\itemsep}{0ex}%
        \setlength{\parsep}{0in}%
        }
    }{\end{list}\par\vspace{12pt}}

\renewenvironment{abstract}{%
    \setlength{\parindent}{0in}%
    \setlength{\parskip}{0in}%
    \bfseries%
    }{\par\vspace{-4pt}}

\newenvironment{addendum}{%
    \setlength{\parindent}{0in}%
    \small%
    \begin{list}{Acknowledgements}{%
        \setlength{\leftmargin}{0in}%
        \setlength{\listparindent}{0in}%
        \setlength{\labelsep}{0em}%
        \setlength{\labelwidth}{0in}%
        \setlength{\itemsep}{12pt}%
        }
    }
    {\end{list}\normalsize}

\usepackage{times}
\usepackage{subfigure}

\usepackage{epstopdf}
\usepackage{graphicx}
\usepackage{epsfig}

\def \aap{Astron.~Astrophys.}
\def \apjl{Astrophys.~J.}
\def \apj{Astrophys.~J.}

\def \mnras{Mon.~Not.~Roy.~Astron.~Soc.}

\def \nat{Nature}

\newcommand{\lsim}{\,\rlap{\raise 0.35ex\hbox{$<$}}{\lower 0.7ex\hbox{$\sim$}}\,}
\newcommand{\gsim}{\,\rlap{\raise 0.35ex\hbox{$>$}}{\lower 0.7ex\hbox{$\sim$}}\,}

\begin{document}

\noindent
 {\Large {\bf {\fontfamily{phv}\selectfont A lower limit of 50 microgauss for the magnetic field near the Galactic Centre}}}

\noindent
Roland M. Crocker,$^{1,2}$ David Jones,$^{2,3,4}$ Fulvio Melia,$^{5}$\\ J{\" u}rgen Ott,$^{6,7}$ \& Raymond J. Protheroe$^{3}$

\begin{affiliations}
 \item J.L. William Fellow, School of Physics, Monash University, Victoria, 3110, Australia
  \item Max-Planck-Institut f{\" u}r Kernphsik, P.O. Box 103980 Heidelberg, Germany
  \item Department of Physics, School of Physics and Chemistry, University of Adelaide, North Terrace,  South Australia, 5005, Australia
  \item Australia Telescope National Facility, Marsfield, 2122, N.S.W., Australia
  \item Physics Department, The Applied Math Program, and Steward Observatory, The University of Arizona, Tucson,  Arizona, AZ 85721, USA
  \item Jansky Fellow, National Radio Astronomical Observatory, Charlottesville, P.O. Box O, 1003 Lopezville Road, Socorro,  New Mexico, NM 87801-0387, USA
  \item California Institute of Technology, 1200 E. California Blvd., Caltech Astronomy, 105-24, Pasadena, CA 91125, USA
\end{affiliations}

\begin{abstract}
The amplitude of the magnetic field near the Galactic Centre has been
uncertain by two orders of magnitude for several decades. On a scale of
$\sim$100 pc fields of $\sim$1000 $\mu$G \cite{Yusef-Zadeh1987,Morris1989,Morris2007} have been reported, 
implying a magnetic energy density more than 10,000 times stronger than
typical for the Galaxy. 
Alternatively, the assumption of
pressure equilibrium between the various phases of the Galactic Centre
interstellar medium (including turbulent molecular gas; the
contested\cite{Revnivtsev2009} ``very hot" plasma;
and the magnetic field) suggests fields of $\sim$100 $\mu$G over $\sim$400
pc size scales\cite{Spergel1992}.
Finally, assuming equipartition, fields of only
$\sim$6 $\mu$G have been inferred from radio observations\cite{LaRosa2005}  for  ~400 pc scales. 
Here we report a compilation of previous
data that reveals a down-break in the region's non-thermal radio spectrum
(attributable to a transition from bremsstrahlung to
synchrotron cooling of the in situ cosmic-ray electron population).
We show that 
the spectral break requires that the Galactic Centre field be
at least $\sim$50
$\mu$G on 400 pc scales, lest the synchrotron-emitting electrons
produce too much $\gamma$-ray emission given existing constraints\cite{Hunter1997}. 
Other considerations support a field of 100 $\mu$G, implying that
$\gsim$10\% of the Galaxy's magnetic energy is contained in only $\lsim$0.05\% of its volume.
%198 words
\end{abstract}

Pinning down the large-scale magnetic field of the Galactic Centre (GC)-- which provides our closest view of a galactic nucleus -- within the greater than two orders of magnitude uncertainty implied by existing, rival analyses has long been of interest.  
A 1000 $\mu$G field would have substantial ramifications for the region's dynamics including
limiting the diffusion distances of relativistic particles (thereby excluding scenarios where the diffuse, $\sim$TeV $\gamma$-ray glow of the GC\cite{Aharonian2006} is ultimately explained as due to a single, astrophysical accelerator\cite{Wommer2008}).
It would also generate a large enough magnetic drag to enhance the in-spiral rate of
giant molecular gas clouds towards the GC\cite{Morris2007}, 
limiting the lifetime of these to $\sim$100 million years, implying `star-bursts' on the same timescale. 
On the other hand, the formation of individual stars might be inhibited (or the stellar initial mass function biased) by a strong magnetic field's  
tendency to support gas against gravitational collapse.

Radio observations at 74 MHz and 330 MHz\cite{LaRosa2005} reveal a diffuse (but distinct) region of non-thermal radio emission
covering the GC (out to $\sim \pm3^\circ$ or $\sim \pm420$ pc from the GC along the Galactic plane).  Invoking
the ``equipartition" condition (minimizing the total energy in magnetic field and relativistic electrons),  a field of only 6 $\mu$G is inferred\cite{LaRosa2005}, typical for the Galaxy-at-large, climbing to 11 $\mu$G over the
inner $\sim \pm0.^\circ8$  ( for extremal parameter values  the field might reach 100 $\mu$G). 

To probe this  radio structure at higher frequencies, 
we have assembled total-intensity, single-dish flux density data at 1.4, 2.4, 2.7, and 10 GHz\cite{Reich1990, Reich1984, Duncan1995,Handa1987}.  
These data are polluted by line-of-sight synchrotron emission 
in the Galactic plane both behind and in front of the GC and the flux density contributed by discrete sources; 
we used a combination of low-pass (spatial wavenumber) filtering to remove the latter
   and all-sky Galactic synchrotron background observations to remove the former
    (see Supplementary Information). After this processing a distinct, non-thermal, radio structure is revealed in all radio maps up to 10 GHz (see ~\ref{fig_10GHzmap}). 
However, the structure's 74 MHz--10 GHz 
spectrum is not described by a pure power-law: attempting to fit 
such to the cleaned data
we find a minimum  $\chi^2$ of 4.9 per degree of freedom ($dof = 4$), excluded at a confidence level of $3.4\sigma$ (\ref{fig_plotChiSqrd}). 

In fact, fitting separate power laws to, respectively, the background-subtracted lower three and upper three radio data, we find that these extrapolations intersect at $\sim$ 1.7 GHz with a spectral break of $\sim$0.6. This is close to the canonical break of 1/2 produced by a steady-state
synchrotron-radiating electron population that transitions (with increasing energy) from 
bremsstrahlung-cooled to synchrotron-cooled. Since the synchrotron cooling rate is a function of magnetic field, $B$, and bremsstrahlung a function of $n_H$,  the break frequency is a function of both parameters.  Observation of a spectral break then determines acceptable pairs of $B$ and $n_H$ for the synchrotron-emitting electrons' environment.

We have modelled the cooled electron distribution and resulting synchrotron emission as a function of $B$, $n_H$, and spectral index at injection of the electron population. We have accounted for losses due to inverse-Compton emission following collisions with ambient light\cite{Porter2006}, ionization, bremsstrahlung and synchrotron emission.
 
The results of our fitting procedure are displayed in  \ref{fig_spctrm} and \ref{fig_plotChiSqrd}. 
A  very important constraint is proffered by $\gamma$-ray data covering the non-thermal emission region:
bremsstrahlung and inverse Compton emission 
will inescapably be generated
by the electrons responsible for the observed radio emission (the energies of electrons synchrotron-radiating at $\sim$GHz and bremsstrahlung-radiating at $\gsim$100 MeV are very similar). This must
not surpass the 300 MeV $\gamma$-ray flux from the region measured by EGRET\cite{Hunter1997} (see  \ref{fig_spctrm}b). 
This consideration rules out field amplitudes $\lsim$ 50 $\mu$G.

In the context of diffuse, $\sim$GeV emission around the GC, we eagerly await results from the Fermi Gamma-ray Space Telescope\cite{Atwood2009} whose sensitivity (more than an order of magnitude better than EGRET's) promises, at worst, an increased lower-limit to the large-scale magnetic field or, more optimistically, a measurement of this field (in concert with radio observations). For now we emphasis that it is a novel analysis -- not new data -- that has allowed the new constraint on the GC magnetic field.
 
\ref{fig_plotEnergyDensity} shows the energy densities of various GC ISM phases.
For acceptable field amplitudes, the cosmic ray electron population is considerably sub-equipartition with respect to the other GC ISM phases (even after accounting for filling factor effects), in particular, the magnetic field, explaining why the magnetic field estimate arrived at 
assuming equipartition\cite{LaRosa2005} is too low. 
We have also calculated the maximum possible energy density in the GC cosmic ray proton population given the EGRET $\gamma$-ray constraints.
Note that at $\sim$ 100 $\mu$G the magnetic field reaches equipartition (at $\sim$300 eV cm$^{-3}$) with the putative `very hot' ($\sim$ 8 keV) phase of the X-ray-emitting plasma supposedly detected throughout the central few degrees along the Galactic plane\cite{Koyama1989}.
Moreover, the energy density of the gas turbulence kinetic energy for the derived $n_H$ is within a factor of a few of equipartition with these other phases up to magnetic field amplitudes of $\sim 100 \  \mu$G.

This situation -- implying near pressure equilibrium between a number of GC interstellar medium phases (including the plasma) -- mirrors that of the Galactic disk, albeit at much higher pressure. Such considerations led 
to the prediction\cite{Spergel1992} that
the real GC magnetic field amplitude lies close to $\sim 100 \ \mu$G.
Very recently, however, observations have shown\cite{Revnivtsev2009} that the X-ray emission 
from a region around $l$ = 0.08, $b$ =1.42  (taken to be typical of the so-called X-ray Ridge) is due to unresolved point sources, implying the very hot X-ray plasma is illusory. 
This casts the pressure equilibrium argument\cite{Spergel1992} into some doubt. Note, however, that
our magnetic field lower limit holds irrespective of whether the 8 keV plasma is real or not.

There are a number of intriguing parallels above to the situation apparently pertaining within the inner regions of starburst galaxies. These independently support the idea that the field is actually close to 100 $\mu$G. 
In starburst environments, it is contended\cite{Thompson2006}, equipartition magnetic field values obtained from radio observations significantly underestimate the real field. Fields are actually sufficiently high to be in hydrostatic equilibrium with the self-gravity of the gaseous disk of the starburst. Such strong fields (together with high gas densities) guarantee that
the relativistic particle population dumps all its energy before being transported out of the system. This calorimetric limit\cite{Voelk1989} may explain\cite{Thompson2006} why even very luminous star-bursting galaxies fall on the far-infrared--radio-continuum correlation\cite{Yun2001}. 
Circumstantial evidence also suggests that radio emission from starbursts is dominated by secondary electrons created in collisions between cosmic ray ions and gas\cite{Thompson2006} (rather than directly-accelerated electrons).

It is noteworthy, then, that  the 60 micron and 1.4 GHz emission from the radio emission region place it within scatter of the far-infrared--radio-continuum correlation\cite{Yun2001}. Furthermore, in the GC
a $\sim$100 $\mu$G field is precisely in the range required to establish hydrostatic equilibrium (given the total and gaseous surface densities). 
A situation wherein a magnetic field provides significant pressure support against gravity may lead to the development of the Parker instability\cite{Parker1966}
and exactly this is suggested by mm-wave observations\cite{Fukui2006} of molecular filaments of several hundred parsec length within $\sim$ 1 kpc of the GC. These observations independently suggest a $\sim$ 100 $\mu$G field.
Finally, the diffuse, $\sim$TeV $\gamma$-ray glow from the vicinity of the GC\cite{Aharonian2006} is most likely explained by cosmic ray impacts with gas. Unavoidably, such collisions would also produce copious secondary electrons which could then contribute significantly to the region's synchrotron radio emission (for fields $\gsim 300 \mu$G, 100\% of the radio emission from the rectangular region shown in  \ref{fig_10GHzmap} could be attributed to secondary electrons).
Taken altogether, these facts paint the Galactic centre as akin to a weak starburst with a magnetic field of $\sim$100 $\mu$G.

%% Here is the endmatter stuff: Supplementary Info, etc.
%% Use \item's to separate, default label is "Acknowledgements"

\begin{addendum}
% \item Put acknowledgements here.
\item[Supplementary Information] is linked to the online version of the paper at www.nature.com/nature. 
 \item[Acknowledgements] RMC thanks Troy Porter for useful conversations about the GC interstellar radiation field.
 DIJ thanks Monash University for hospitality. RMC and DIJ thank John Dickey for advice about radio data analysis.
\item[Author Contributions] R.M.C. led the work and performed the main analysis. D.I.J. performed the analysis of radio data including development of the Fourier-based technique for background and foreground removal, was responsible for original radio observations, and provided critical scientific discussion. F.M. provided input on theoretical and statistical problems and critical discussion of scientific interpretation. J.O. supervised the analysis of archival radio data and the taking of original radio data and provided input on statistics. R.J.P. provided input on thermal and relevant non-thermal processes
and critical discussion of scientific interpretation.  R.J.P.  and R.M.C. provided supervision of D.I.J. as doctoral candidate. All the authors discussed the results and commented on the manuscript.  
 \item[Author Information] The authors declare that they have no
competing financial interests. Correspondence and requests for materials
should be addressed to R.M.C.~(email: Roland.Crocker@mpi-hd.mpg.de). 
Reprints and permissions information is available at npg.nature.com/reprintsandpermissions.
 %\item[Correspondence] 
\end{addendum}

\pagebreak

\renewcommand{\thesection}{Figure \arabic{section}}

\begin{figure}
\renewcommand \thefigure{1}
\centerline{\includegraphics[width=\textwidth]{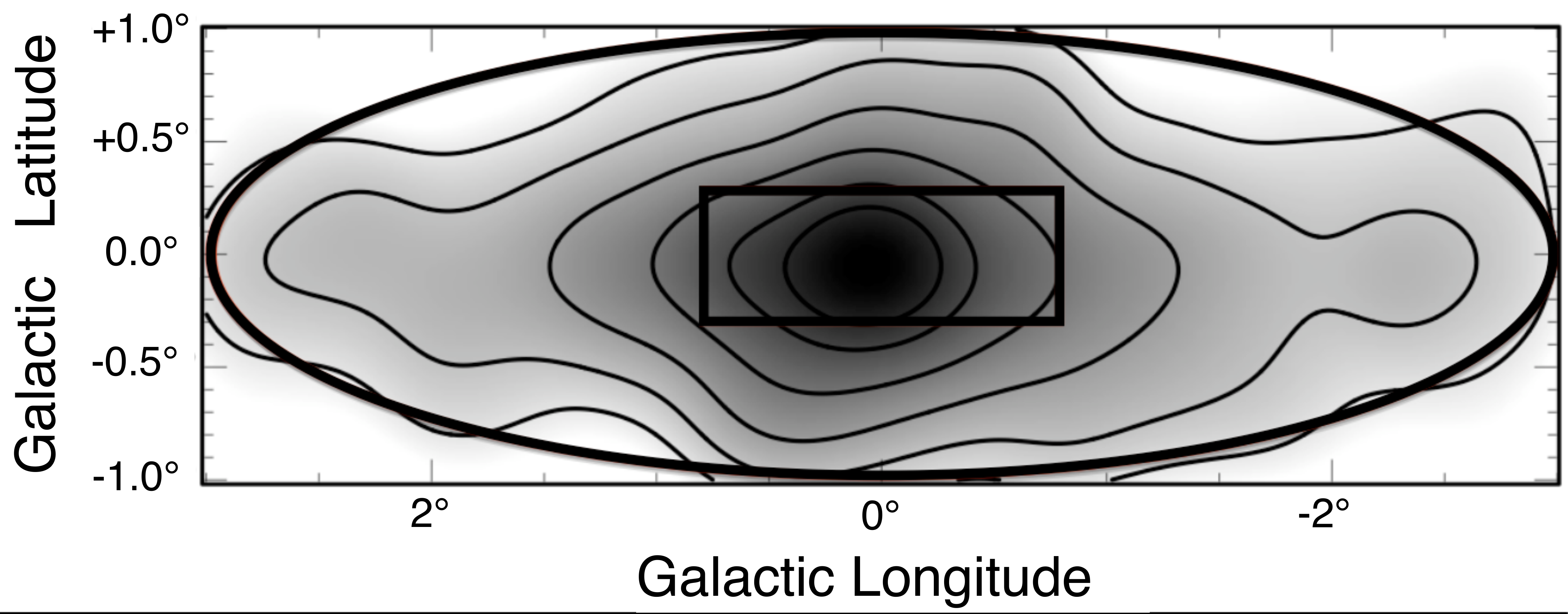}}
\end{figure}

\noindent
\section{}
\label{fig_10GHzmap}
{\bf Figure 1 $|$ Total intensity image
of the region at 10 GHz}. Radio map\cite{Handa1987} convolved to a resolution of 1$^\circ.2 \times 1^\circ$.2  with contours at  
10, 20, 40, 80, 160 and 240 Jy/beam. (Native resolution and convolved images at $\nu \geq$ 1.4 GHz in the Supplementary Information.)
There is a striking constancy in the appearance of the radio structure from 74 MHz to at least 10 GHz (the large ellipse traces the diffuse, non-thermal radio emission region first identified at 74 and 330 MHz\cite{LaRosa2005}).
The small  rectangle delineates the region from which the HESS collaboration determines a diffuse $\sim$ TeV $\gamma$-ray intensity\cite{Aharonian2006}.
%94 words

\pagebreak

\begin{figure*}[h]
\renewcommand \thefigure{2}
\subfigure[]{
\includegraphics[width=0.5\columnwidth]{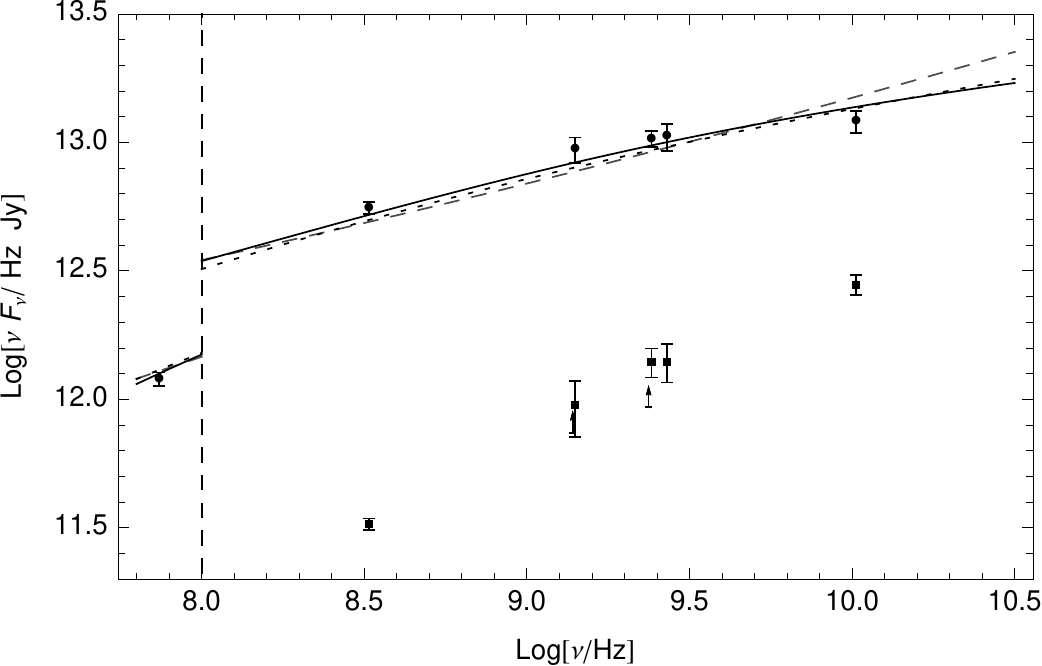}
} 
\subfigure[]{
\includegraphics[width=0.5\columnwidth] {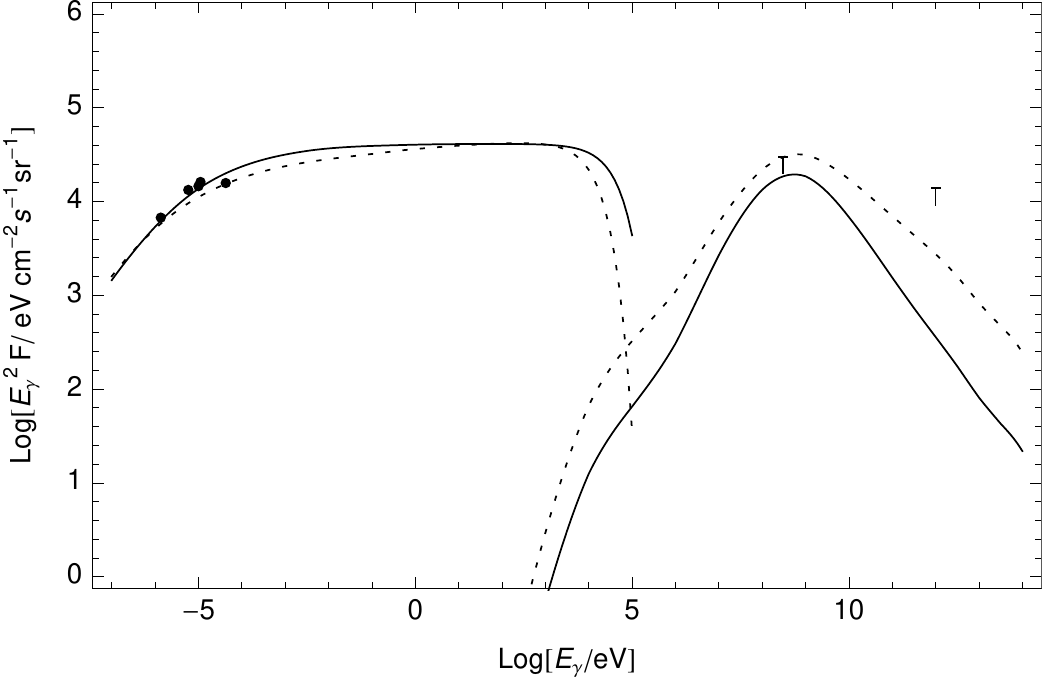}
} 
\caption{}
\end{figure*}

\noindent
\section{}
\label{fig_spctrm}
{\bf Figure 2 $|$ Spectrum of the region: data and models.}
({\bf a}) {\bf Radio}. 
Points:--{\bf circles}: flux density of the radio structure 
after removal of flux from sources with sizes $< 1^\circ.2$.
The vertical line divides the 74 MHz datum (that
receives no contribution from Galactic plane synchrotron background/foreground) from all the other data (that do) -- hence the discontinuity in the model curves at 100 MHz;
{\bf squares}: discrete source flux densities (measured directly by the VLA at 330 MHz[\cite{LaRosa2005}]  and otherwise obtained through the Fourier technique described in the Supplementary Information);
{\bf arrows}: discrete source lower limits obtained from ATCA. The error bars show 68\% confidence intervals. 
{\bf Curves}: the best-fit flux density due to synchrotron emission from a cooled electron distribution plus Galactic plane synchrotron  for the case of  ({\bf solid}) 100 $\mu$G and ({\bf dotted}) 30 $\mu$G fields 
(the break in the radio synchrotron curve mirrors a corresponding break in the distribution of radiating electrons; we assume the electrons are injected with a power law in momentum\cite{Crocker2007}); 
{\bf dashed}:  the best-fit null hypothesis case of a pure power-law signal plus Galactic plane synchrotron.
({\bf b}) {\bf Broadband.}
Models for magnetic field of 100 $\mu$G (solid) and 30 $\mu$G (dotted). 
The injected electron spectrum is assumed to follow a power law up to $\sim 100$ TeV. 
(Note that the cut-off energy is only weakly constrained.
Our conclusions are not sensitive to the exact cut-off energy, however, given that radio observations guarantee
the existence of $\sim$GeV electrons whose bremsstrahlung emission we can compare against the EGRET intensity datum.)
The lower energy curves are synchrotron emission. The higher are bremsstrahlung plus inverse Compton emission from the same electrons responsible for the synchrotron (a figure showing the individual bremsstrahlung and inverse Compton contributions is shown in the Supplementary Information). The upper limits are due to observations by EGRET\cite{Hunter1997} at 300 MeV and HESS\cite{Aharonian2006} at TeV. 
%300 words

\pagebreak

\begin{figure}
\renewcommand \thefigure{4}
\centerline{\includegraphics[width=\textwidth]{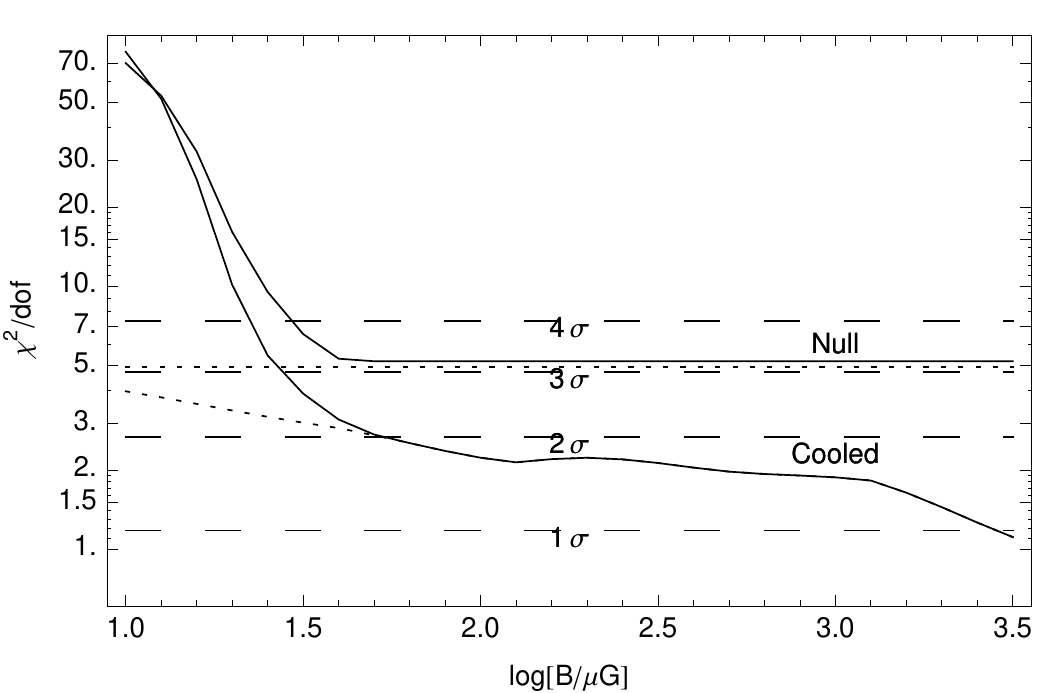}}
\end{figure}

\noindent
\section{}
{\bf Figure 3 $|$ Plot of the $\chi^2$ per degree of freedom as a function of magnetic field amplitude.}
\label{fig_plotChiSqrd}
Curves: (as labelled) model for a cooled primary electron model (with 3 degrees of freedom) and the null hypothesis of a pure power law electron distribution (with 4 $dof$). The solid curves are constrained by the requirement that the $\gamma$-ray emission from the synchrotron-radiating electron population be less than
the  upper limit obtained from EGRET data\cite{Hunter1997} (dotted curves not so constrained).
The horizontal dashed lines mark the 1,2,3,4$\sigma$ confidence limits for a model with 3 $dof$ and do not apply to the null hypothesis (which only achieves a best fit acceptable at the $\sim 3.4 \sigma$ level).
%107 words

\pagebreak

\begin{figure}
\renewcommand \thefigure{5}
\centerline{\includegraphics[width=\textwidth]{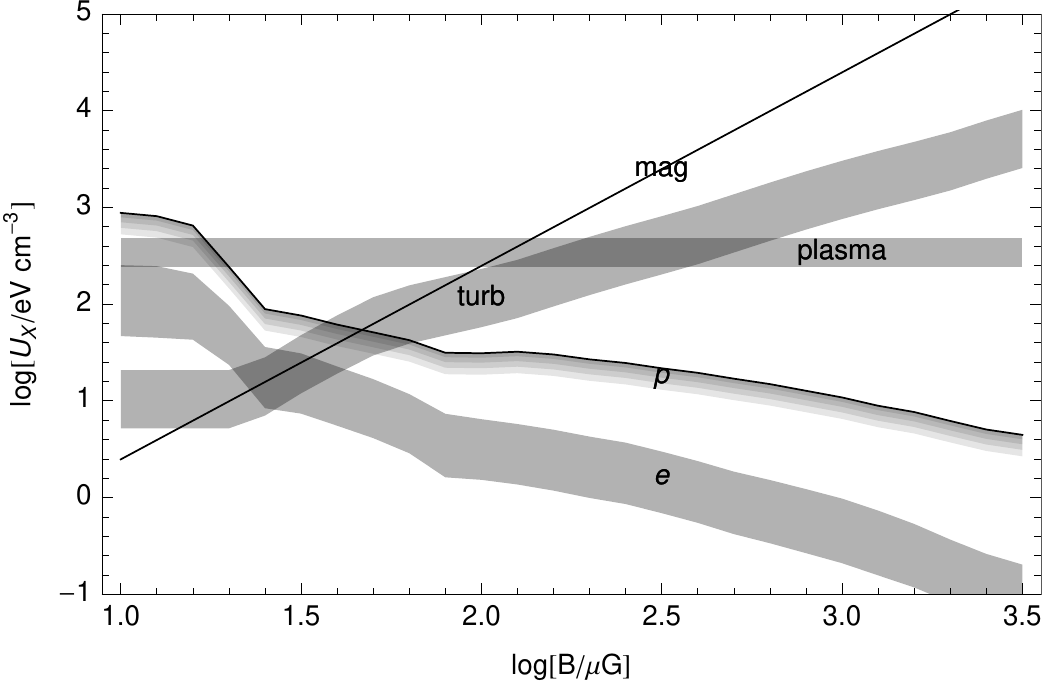}}
\end{figure}

\noindent
\section{}
{\bf Figure 4 $|$ Energy density in various phases of the GC interstellar medium as a function of magnetic field.} 
\label{fig_plotEnergyDensity}
Bands show 68\% confidence limits (except for $p$).
Labels denote:-- `mag': magnetic field; `plasma': the disputed\cite{Revnivtsev2009} X-ray emitting 8 keV plasma (with a density\cite{Yamauchi1990} 0.03--0.06 cm$^{-3}$); `turb': turbulent motions of the local gas (assuming it is at the best-fit $n_H$ and has a velocity dispersion in the range 15--30 km/s typical for GC molecular clouds\cite{Gusten2004}  -- note that, as the $n_H$ required to fit the radio spectrum increases with $B$, the turbulent energy density is an increasing function of $B$); `$e$': cosmic ray electrons; 
and `$p$':  a conservative upper bound on the cosmic ray proton energy density inferred from the 300 MeV $\gamma$-ray upper limit (a putative proton population, colliding with ambient gas at the best-fit $n_H$, would produce $\gamma$-rays mostly via neutral meson decay).
The electron and proton energy densities take into account the filling factor of gas at the given value of $n_H$[\cite{Paglione1998}] (we take the filling factor of gas within $1\sigma$ around the best-fit $n_H$ value); this is a correction upwards to the values of $U_e$ and $U_p$ by an amount 100--1000. Without this correction, one would have $U_e \simeq U_B$ at $\sim$ 10 $\mu$G in agreement with the estimate derived assuming equipartion\cite{LaRosa2005}. A factor of 2 uncertainty\cite{Ferriere2007} in the hydrogen mass of the region is also accounted for in determining the electron and proton energy densities.
%226 words

\end{document}